\documentclass{article}

\usepackage{microtype}
\usepackage{graphicx}
\usepackage{subcaption}
\usepackage{booktabs} % for professional tables
\usepackage{xurl}
\usepackage{hyperref}
\usepackage{wrapfig}

\usepackage[preprint]{icml2026}

\usepackage{amsmath}
\usepackage{amssymb}
\usepackage{mathtools}
\usepackage{amsthm}
\usepackage{tabularx}
\usepackage{arydshln}
\usepackage{booktabs}
\usepackage{multirow}

\usepackage[capitalize,noabbrev]{cleveref}

\theoremstyle{plain}

\theoremstyle{definition}

\theoremstyle{remark}

\usepackage{array}
\usepackage{pifont}

\newcolumntype{C}[1]{>{\centering\arraybackslash}p{#1}}

\usepackage[textsize=tiny]{todonotes}

\icmltitlerunning{AudioEval: Automatic Dual-Perspective and Multi-Dimensional TTA Evaluation}

\begin{document}

\twocolumn[
  \icmltitle{AudioEval: Automatic Dual-Perspective and Multi-Dimensional Evaluation of Text-to-Audio-Generation}



  \icmlsetsymbol{equal}{*}

  \begin{icmlauthorlist}
    \icmlauthor{Hui Wang}{yyy}
    \icmlauthor{Jinghua Zhao}{yyy}
    \icmlauthor{Junyang Chen}{yyy}
    \icmlauthor{Cheng Liu}{yyy}
    \icmlauthor{Yuhang Jia}{yyy}
    \icmlauthor{Haoqin Sun}{yyy}
    \icmlauthor{Jiaming Zhou}{yyy}
    \icmlauthor{Yong Qin}{yyy}
  \end{icmlauthorlist}

  \icmlaffiliation{yyy}{College of Computer Science, Nankai University, Tianjin, China}

  \icmlcorrespondingauthor{Yong Qin}{qinyong@nankai.edu.cn}

  \icmlkeywords{Text-to-Audio Generation, Auio Quality Assessment}

  \vskip 0.3in
]



\printAffiliationsAndNotice{}  % no special notice (required even if empty)

\begin{abstract}
  Text-to-audio (TTA) generation is advancing rapidly, but evaluation remains challenging because human listening studies are expensive and existing automatic metrics capture only limited aspects of perceptual quality. We introduce AudioEval, a large-scale TTA evaluation dataset with 4,200 generated audio samples (11.7 hours) from 24 systems and 126,000 ratings collected from both experts and non-experts across five dimensions: enjoyment, usefulness, complexity, quality, and text alignment. Using AudioEval, we benchmark diverse automatic evaluators to compare perspective- and dimension-level differences across model families. We also propose Qwen-DisQA as a strong reference baseline: it jointly processes prompts and generated audio to predict multi-dimensional ratings for both annotator groups, modeling rater disagreement via distributional prediction and achieving strong performance. We will release AudioEval to support future research in TTA evaluation.
\end{abstract}

\section{Introduction}
% TTA系统的快速发展,广泛的前景
In recent years, text-to-audio (TTA) technology has emerged as an important and rapidly evolving research area at the intersection of natural language processing and audio generation~\citep{huang2023make1, huang2023make2, tango, liu2023audioldm}. Unlike conventional text-to-speech (TTS) systems~\citep{valle, felle} that focus on naturalness and intelligibility, TTA aims to generate diverse audio content from text, extending text-conditioned audio generation beyond speech. This broader scope enables richer multimodal interaction and supports applications in virtual reality, accessibility, and creative media. However, the open-ended nature of TTA makes evaluation challenging, which limits the ability to benchmark systems and hinders further advancement in the field.

Current evaluation practices of TTA typically combine subjective and objective approaches. Subjective evaluation mainly relies on human ratings, commonly reported as Mean Opinion Scores (MOS). While human judgment is considered the gold standard, it is expensive and time-consuming~\citep{wang23r_interspeech}. On the other hand, objective metrics from related domains, such as Frechet Inception Distance~\citep{fid} and CLAP~\citep{CLAP2023}, have been adopted for automatic evaluation~\cite{fad}. Although these metrics offer efficiency, they provide a limited view and often fail to align with human perception~\cite{fad-correlates-poorly}. Some also require reference audio, which restricts their applicability. Overall, effective and reliable evaluation tools tailored to the characteristics of TTA remain lacking.

% challenge 1: 缺乏可用的数据，过去关注的对象都不行
Automatic perceptual evaluation has demonstrated effectiveness in TTS, voice conversion (VC), and text-to-music (TTM), offering both efficiency and consistency with human perception~\cite{utmos, aes, musiceval, songeval}. These developments suggest its potential for advancing TTA evaluation. However, a key challenge in applying such techniques to TTA lies in the lack of standardized, large-scale human-annotated evaluation datasets. Existing datasets are often built on narrow domains such as English speech, music, or singing, which limits their generalizability to open-domain audio. Human evaluations specifically designed for TTA are also limited in scale and coverage, are not always publicly available, and often follow inconsistent protocols, further hindering progress in this direction.

\begin{table*}[t]
\caption{Comparison of human-annotated datasets for audio quality assessment. 
``View-separated'' indicates whether annotations are collected with explicitly separated rater views. 
TA denotes whether the dataset supports \textit{text--audio alignment} evaluation.}
\label{tab:dataset_comparison}
\centering
\small
\setlength{\tabcolsep}{4pt}
\renewcommand{\arraystretch}{1.12}

\begin{tabular}{l l c cc c c c}
\toprule
\textbf{Dataset} & \textbf{Source of Data} &
\textbf{View-separated} &
\multicolumn{2}{c}{\textbf{Eval. dimensions}} &
\multicolumn{3}{c}{\textbf{Annotation scale}} \\
\cmidrule(lr){4-5}\cmidrule(lr){6-8}
& & & \textbf{Dim} & \textbf{TA} &
\textbf{Clips} & \textbf{Raters/clip} & \textbf{Ratings} \\
\midrule
BVCC~\citep{bvcc} & Speech (TTS/VC) &
$\times$ &
1 & $\times$ &
7{,}106 & 8 & 56{,}848 \\

MusicEval~\citep{musiceval} & Music (TTM) &
$\times$ &
2 & \checkmark &
2{,}748 & 5 & 27{,}480 \\

AES-Natural~\citep{aes} & Real Audio &
$\times$ &
4 & $\times$ &
2{,}950 & 10 & 118{,}000 \\

SongEval~\citep{songeval} & Song &
$\times$ &
5 & $\times$ &
2{,}399 & 4 & 47{,}980 \\

\textbf{AudioEval (Ours)} & \textbf{General audio (TTA)} &
\checkmark &
\textbf{5} & \checkmark &
\textbf{4{,}200} & \textbf{6 (3+3)} & \textbf{126{,}000} \\
\bottomrule
\end{tabular}
\end{table*}

% challenge 2：只支持单一视角、受限维度的评估
Another key challenge is that existing automatic evaluation methods typically focus on a limited set of quality dimensions; moreover, they fail to differentiate between evaluation groups, leading to differences in perspective being overlooked. TTA outputs are inherently multi-faceted, with quality varying along aspects such as enjoyability, usefulness, complexity, production quality, and textual alignment. These aspects are especially important when TTA systems are applied in different downstream scenarios. Moreover, different user groups such as experts and lay users often interpret these aspects differently~\citep{10.1145/3769106}. A meaningful evaluation tool should therefore support multi-aspect and multi-perspective assessment to enable accurate diagnosis, fair comparison, and practical deployment.

To address these challenges, we introduce AudioEval. As far as we know, it is the first dataset for evaluation of TTA-generated audio, enabling automated, dual-perspective, and multi-dimensional assessment. It includes 4,200 audio samples with 25,200 records and 126,000 dimension-level ratings. Both experts and non-experts contribute, capturing complementary perspectives of audio perception. Using AudioEval, we evaluate a range of automatic evaluators for TTA quality prediction, studying how well different model families align with expert and non-expert judgments across dimensions. As part of this evaluation, we include Qwen-DisQA, an automatic quality scoring model based on Qwen2.5-Omni \cite{qwen25omni}. It jointly processes textual prompts and generated audio to predict multi-dimensional ratings from both expert and non-expert perspectives, and models rater disagreement via distributional prediction to provide more nuanced evaluations.

% 贡献：数据集+模型+实验
In summary, our contributions are three-fold:
\begin{itemize}
    \item We present \textbf{AudioEval}, the first multi-dimensional TTA evaluation dataset with ratings from both experts and non-experts, supporting automated evaluation.
    \item We develop \textbf{Qwen-DisQA} as a reference automatic scoring model, which predicts perceptual ratings from text--audio pairs and captures rater disagreement through distribution modeling.
    \item We conduct experiments to evaluate \textbf{diverse automatic evaluators} on AudioEval for TTA quality prediction, highlighting their strengths and limitations across dimensions and annotator perspectives.
\end{itemize}

\section{Related Work}

\subsection{TTA Systems and Evaluation}
Text-to-audio has rapidly progressed from early text-conditioned waveform generation toward general-purpose, open-domain audio synthesis. Recent representative systems are largely built upon latent diffusion frameworks, where a text encoder conditions a generative model operating in a compact audio representation space. For example, AudioLDM and its variants demonstrate the feasibility of synthesizing diverse sound events from natural-language prompts, bridging text understanding and audio generation in a unified pipeline~\citep{liu2023audioldm}. In parallel, several works focus on improving generation fidelity, controllability, and coverage, including Make-An-Audio series~\citep{huang2023make1, huang2023make2} and text-to-audio generation via staged or compositional training objectives~\citep{tango}.

Evaluation for text-to-audio is often inherited from neighboring domains and typically combines subjective listening tests with automatic proxy metrics. Human evaluations, such as MOS-style ratings or pairwise preferences, best reflect perceived quality, but they are costly and sensitive to rater expertise, task instructions, and prompt selection~\citep{chiang23_interspeech}. Automatic evaluation commonly relies on distributional similarity measures~\citep{fid,fad} and audio–text alignment scores from contrastive models such as CLAP~\citep{CLAP2023}. However, these proxies are incomplete: alignment metrics can miss perceptual artifacts, while distributional distances may correlate weakly with human judgments, especially under diverse prompts or when reference sets are mismatched~\citep{fad-correlates-poorly}. These limitations motivate dedicated protocols and human-annotated benchmarks that better capture the multi-dimensional nature of TTA outputs.

\subsection{Automatic Perceptual Quality Prediction}
Automatic perceptual quality prediction learns models to approximate human judgments, providing scalable alternatives to expensive listening tests. In TTS and voice conversion, MOS predictors trained on large-scale ratings have shown strong correlation with subjective evaluation, making them useful for rapid benchmarking~\citep{utmos}. Recent work further argues that perceptual quality is multi-faceted and benefits from multi-dimensional protocols; for example, AES-style frameworks annotate general audio along multiple criteria to support more diagnostic evaluation~\citep{aes}. Similar multi-aspect trends also appear in music and singing generation evaluation~\citep{musiceval, songeval}.

From the perspective of open-domain TTA, prior resources remain limited. As summarized in Table~\ref{tab:dataset_comparison}, most existing datasets are built on narrow domains (speech, music, or song), typically offer fewer evaluation dimensions, and often do not explicitly support text--audio alignment assessment, which is central to prompt-conditioned generation. Moreover, annotations are usually collected from a single rater pool without separating perspectives (e.g., experts vs.\ non-experts), potentially overlooking systematic differences in criteria. These gaps motivate perceptual predictors that jointly model text--audio pairs and provide multi-dimensional, view-aware scoring tailored to TTA.

\section{AudioEval Dataset}
\label{sec:format}

AudioEval is a dataset for evaluating text-to-audio generation from both expert and non-expert perspectives across five key dimensions. This section is organized into three parts: data collection, annotation, and analysis.

\begin{table}[t]
\centering
\caption{Statistics of the 24 TTA systems included in AudioEval.}
\label{tab:year_and_size}
\small

\begin{subtable}{\linewidth}
\centering
\begin{tabular*}{0.9\linewidth}{@{\extracolsep{\fill}}lccccc@{}}
\toprule
Year & Pre-2022 & 2022 & 2023 & 2024 & 2025 \\
\midrule
\#Sys. & 0 & 1 & 6 & 12 & 5 \\
\bottomrule
\end{tabular*}
\end{subtable}

\medskip

\begin{subtable}{\linewidth}
\centering
\begin{tabular*}{0.9\linewidth}{@{\extracolsep{\fill}}lccccc@{}}
\toprule
Size & {$<$0.5B} & {[0.5,1)B} & {[1,2)B} & {$\ge$2B} & {Unknown} \\
\midrule
\#Sys. & 3 & 8 & 6 & 1 & 6 \\
\bottomrule
\end{tabular*}
\end{subtable}

\end{table}

\begin{figure}
    \centering
    \includegraphics[width=1\linewidth]{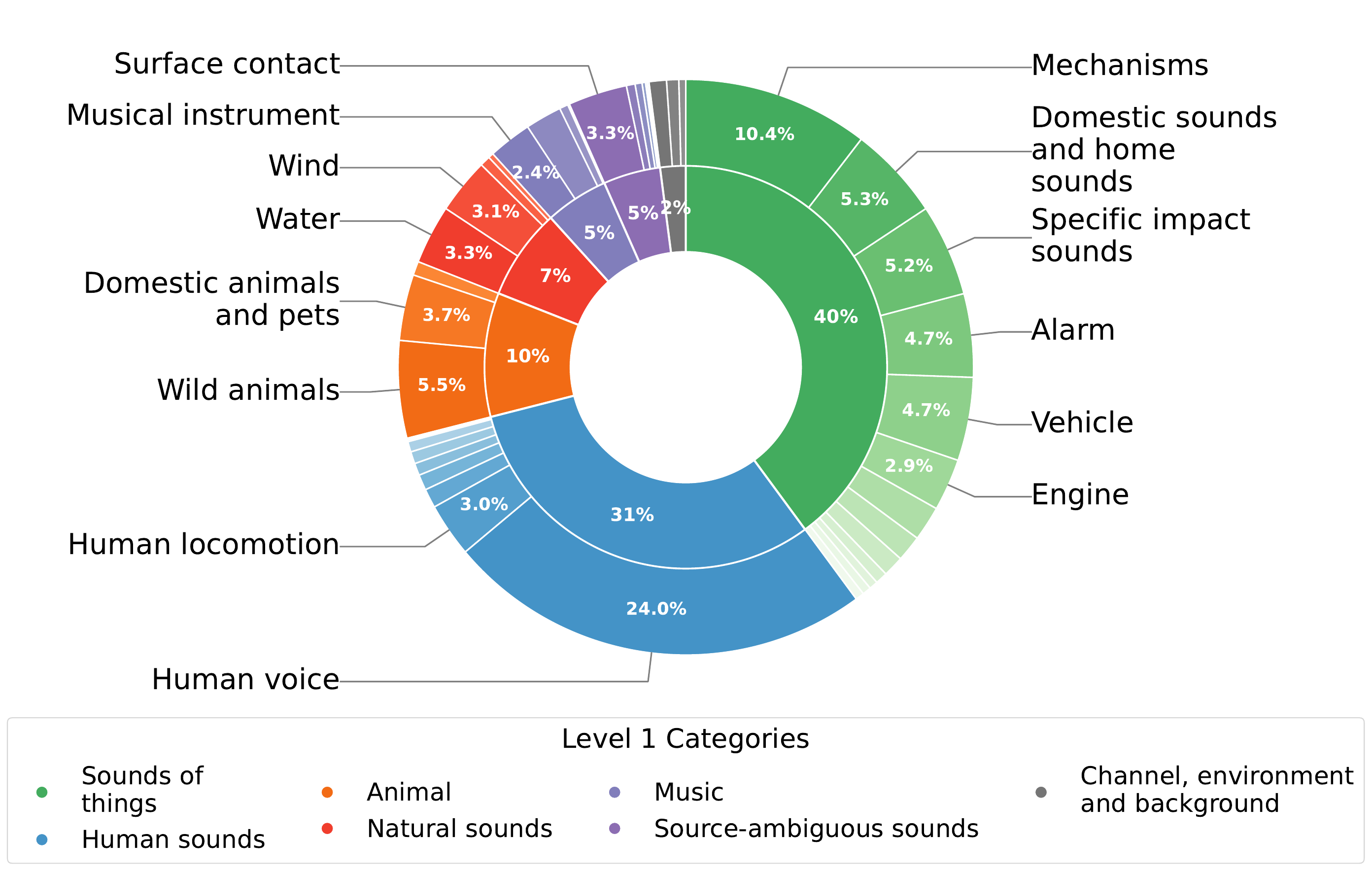}
    \caption{Distribution of AudioEval prompts over sound event categories based on the AudioSet ontology. The inner ring shows top-level groups; the outer ring shows subcategories.}
    \label{fig:prompt_audioset}
\end{figure}

\begin{figure}[t]
  \centering
  \centerline{\includegraphics[width=\linewidth]{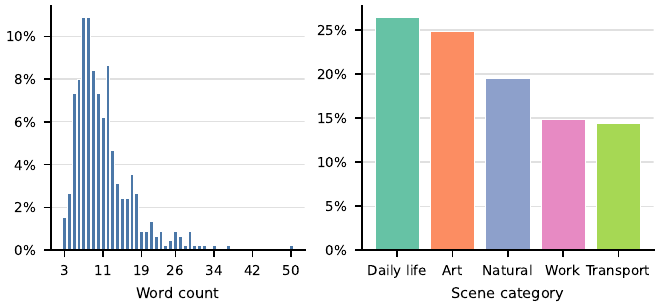}}
 
  \caption{Prompt-level diversity in AudioEval. Left: Distribution of prompt lengths. Right: Distribution across scene types.}
  \label{fig:prompt_diversity}
\end{figure}

\subsection{Data Collection}

AudioEval collects 4,200 text-conditioned audio generations (11.7h total) from 24 TTA systems and a curated prompt set that covers a wide range of acoustic scenes, sound sources, and descriptive complexity, with audio sourced from both locally executed inference and public demo interfaces.

\paragraph{Systems.} 

AudioEval covers 24 representative TTA systems to reflect the diversity of modern generators and enable cross-system comparison. The systems span major modeling paradigms, including autoregressive approaches (e.g., AudioGen~\cite{audiogen}), diffusion and latent-diffusion families (e.g., AudioLDM series~\cite{liu2023audioldm, audioldm2-2024taslp}, Make-An-Audio~\cite{huang2023make1, huang2023make2}, Tango~\cite{tango,tango2}), and accelerated generation methods such as consistency/LCM-style models (e.g., ConsistencyTTA~\cite{consistencytta}, AudioLCM~\cite{audiolcm}, SoundCTM~\cite{soundctm}). As shown in Table~\ref{tab:year_and_size}, the collection is skewed toward recent systems (especially 2024--2025) and covers a wide range of parameter scales, which helps analyze how model recency and size relate to multi-dimensional quality. The full list of systems is provided in Appendix~\ref{app:system_list}. 

% 类别图
\paragraph{Prompt.} 

AudioEval includes 451 prompts designed to cover a wide range of sound events and real-world scenarios, while varying in linguistic complexity and specificity. Figure~\ref{fig:prompt_audioset} analyzes the sound event diversity based on the AudioSet ontology~\citep{audioset}. The outer ring reflects the distribution over second-level sound categories, while the inner ring shows the broader top-level groups such as human sounds, music, animals, and environmental sounds. The coverage is well-balanced, which ensures that TTA outputs span a broad and realistic spectrum of auditory content. Figure~\ref{fig:prompt_diversity} presents two complementary analyses at the sentence level. The left histogram shows the word count distribution across prompts, with most falling in the 5–20 word range, reflecting realistic descriptive lengths while including both concise and detailed inputs. The right bar chart categorizes prompts into five scene types based on the TTA-Bench taxonomy~\citep{ttabench}. The prompt set is dominated by Daily Life and Art scenes, but includes meaningful coverage of all categories, supporting evaluation across diverse usage contexts. Other prompt information is provided in Appendix~\ref{app:prompt}.

\begin{figure*}[ht]
  \centering
  \centerline{\includegraphics[width=\linewidth]{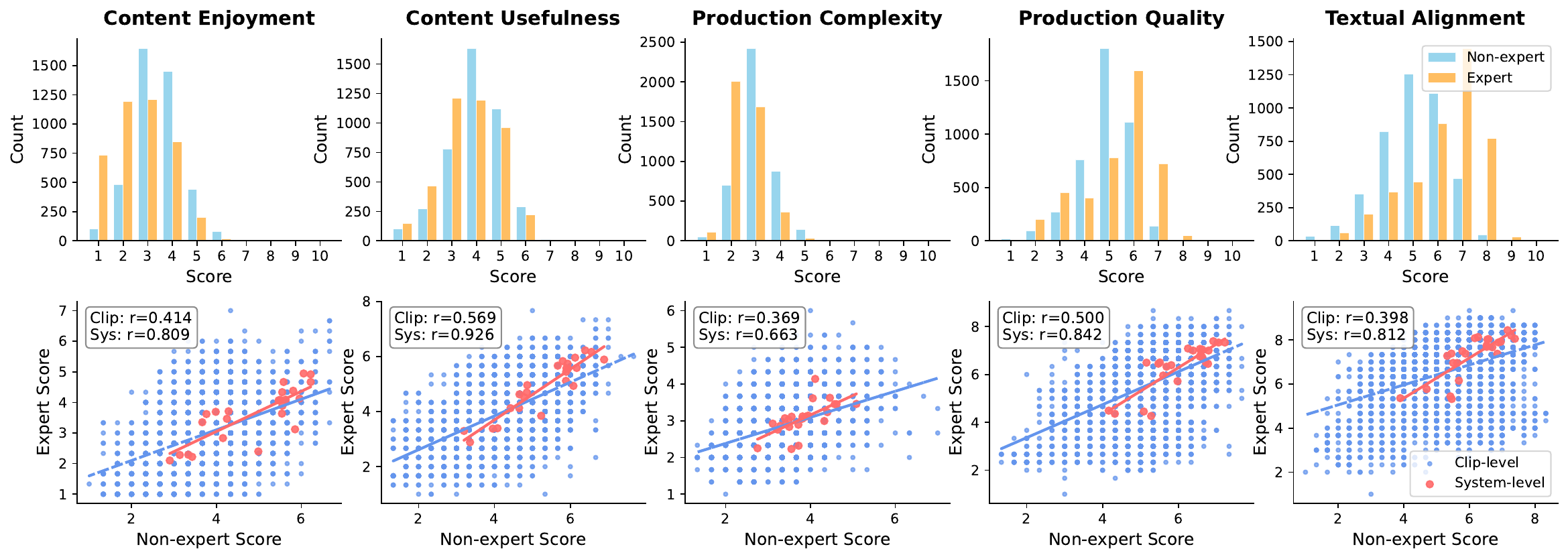}}
  \caption{Top: score distributions of expert and non-expert raters across five evaluation dimensions. Bottom: correlations between expert and non-expert scores at the clip level (individual utterances) and the system level (per-system averages).}
  \label{fig:distribution_comparison}
\end{figure*}

\begin{table}[t]
  \caption{Five dimensions for evaluation in AudioEval.}
  \vspace{-4pt}
  \label{tab:metric}
  \begin{center}
    \begin{small}
      \setlength{\tabcolsep}{4pt}        % 列间距
      \renewcommand{\arraystretch}{1.05} % 行高（可选）

      \begin{tabularx}{\linewidth}{
        @{\hspace{5pt}}                  % 左侧留白（关键）
        >{\raggedright\arraybackslash}p{0.23\linewidth}
        >{\raggedright\arraybackslash}X
        @{\hspace{4pt}}                  % 右侧留白（关键）
      }
        \toprule
        \textbf{Dimension} & \textbf{Definition} \\
        \midrule

        Content Enjoyment & Degree of subjective enjoyment, including emotional impact and artistic expression. \\
        \addlinespace[2pt]\hdashline[0.8pt/2pt]\addlinespace[2pt]

        Content Usefulness & Potential usefulness of the audio for downstream applications or creative purposes. \\
        \addlinespace[2pt]\hdashline[0.8pt/2pt]\addlinespace[2pt]

        Production Complexity & Level of acoustic richness and diversity of structural elements. \\
        \addlinespace[2pt]\hdashline[0.8pt/2pt]\addlinespace[2pt]

        Production Quality & Technical fidelity of the audio, covering clarity, dynamics, and balance. \\
        \addlinespace[2pt]\hdashline[0.8pt/2pt]\addlinespace[2pt]

        Textual Alignment & Accuracy of semantic and temporal alignment with the input text. \\

        \bottomrule
      \end{tabularx}
    \end{small}
  \end{center}
  \vskip -0.1in
\end{table}

\begin{figure}[t]
  \centering
  \centerline{\includegraphics[width=0.6\linewidth]{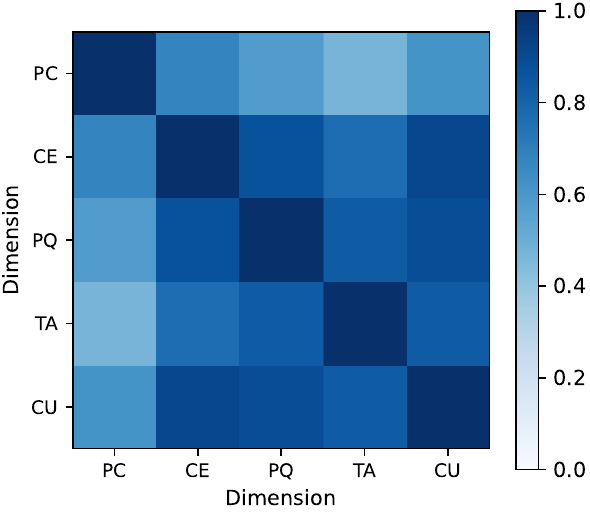}}
 
  \caption{Pearson correlation matrix among the five evaluation dimensions, computed across all annotated samples.}
  \label{fig:dim_analysis}
\end{figure}

\subsection{Annotation Protocol}
\label{sec:annotation_protocol}

\paragraph{Annotators.}
AudioEval adopts a two-population design to capture both professional judgment and end-user perception. These two groups are defined as follows.
\begin{itemize}
\item \textbf{Experts}, with academic training in audio engineering, speech, or music, who provide reliable references based on professional judgment.
\item \textbf{Non-experts}, recruited from a general listener population, who provide user-centered impressions relevant for real-world applications.
\end{itemize}

Each audio sample receives independent scores from three expert and three non-expert annotators. Our annotator pool includes 3 expert annotators and 9 non-expert annotators, all with sufficient English proficiency (CET-4 or above) to comprehend the prompts. Expert annotators hold higher education degrees in music-related fields and bring substantial experience in listening and evaluating sound, with an average age of 40.7. Non-expert annotators come from a variety of non-audio academic backgrounds. Detailed information about the annotators can be found in Appendix~\ref{app:rater}.

\paragraph{Evaluation Dimensions.}

Inspired by prior work, we adopt a five-dimensional evaluation framework, including Content Enjoyment (CE), Content Usefulness (CU), Production Complexity (PC), Production Quality (PQ), and Textual Alignment (TA), as summarized in Table~\ref{tab:metric}.

Each dimension captures a distinct aspect of quality. CE and CU focus on perceptual experience and functional value, while PC and PQ assess technical characteristics such as acoustic richness and fidelity. TA measures how accurately the audio reflects the prompt. These dimensions are complementary; for example, a sample may align well with the text but still lack production quality or user appeal. Using multiple dimensions helps identify specific strengths and weaknesses of TTA systems and supports fine-grained supervision for training evaluation models. It also enables better diagnostic analysis and more targeted comparison across different models. The detailed scoring criteria for each dimension are provided in Appendix~\ref{app:scoring_guidelines}.

\paragraph{Annotation Process.}

All annotators receive written guidelines that define each evaluation dimension and provide example cases, followed by a brief calibration phase to standardize interpretation. Annotation is conducted using a web-based interface that presents the text prompt, embedded audio playback, and five 10-point scoring sliders corresponding to the evaluation dimensions. Annotators are allowed to replay the audio as needed. The interface hides system identities and randomizes sample presentation to reduce bias. Each audio sample is independently rated by three expert and three non-expert annotators. Annotators are instructed to evaluate each dimension separately rather than provide a holistic judgment. To minimize fatigue, the annotation process is divided into multiple sessions, with each session limited to a maximum of 50 audio samples. All scores are aggregated and aligned by sample, group, and dimension for downstream analysis.

\paragraph{Quality Control.}

We implement multiple measures to ensure the consistency and reliability of annotations. All annotators must complete a training and qualification phase before contributing to the dataset, during which their understanding of the evaluation dimensions and use of the scoring scale is assessed. During annotation, we insert probe samples to evaluate internal consistency: for a small subset of items, the same audio sample is presented twice, and if an annotator gives scores differing by more than two points on the same dimension, both responses are discarded. In addition, we manually inspect samples with unusually high score variance across annotators, and flag any systematic patterns of inattentive or biased behavior. These procedures help maintain score quality across sessions and annotator groups, and ensure the dataset reflects stable human judgments.

\subsection{Dataset Analysis}

\paragraph{Comparison Between Expert and Non-Expert Perspectives}

We analyze the annotated scores to understand both differences between annotator groups and the structure among evaluation dimensions. As shown in Figure~\ref{fig:distribution_comparison} (top), non-expert annotators generally assign higher scores than experts, especially in Content Enjoyment, Content Usefulness, and Textual Alignment. Experts apply more conservative judgments, particularly in technical and semantic aspects. Despite these differences, Figure~\ref{fig:distribution_comparison} (bottom) shows strong agreement at the system level, with Pearson correlations exceeding 0.66 across all dimensions. At the clip level, the agreement is moderate (ranging from 0.37 to 0.57), indicating higher variability in individual judgments.

\paragraph{Analysis of Inter-Dimensional Relationships}
Figure~\ref{fig:distribution_comparison} (top) also reveals distinct score distribution patterns across dimensions. Content Enjoyment and Content Usefulness are skewed toward higher ratings, suggesting that most samples achieve a baseline level of perceptual appeal. Production Complexity shows a wider spread, while Production Quality and Textual Alignment vary significantly across systems. Figure~\ref{fig:dim_analysis} presents the correlation matrix among dimensions, where we observe strong associations between Content Enjoyment and Content Usefulness, and between Production Quality and Textual Alignment. In contrast, Production Complexity is relatively independent, highlighting its role in capturing structural richness rather than surface-level fidelity. These patterns confirm the value of incorporating multiple dimensions and perspectives for comprehensive quality assessment.

\begin{figure}[t]
  \centering
  \centerline{\includegraphics[width=\columnwidth]{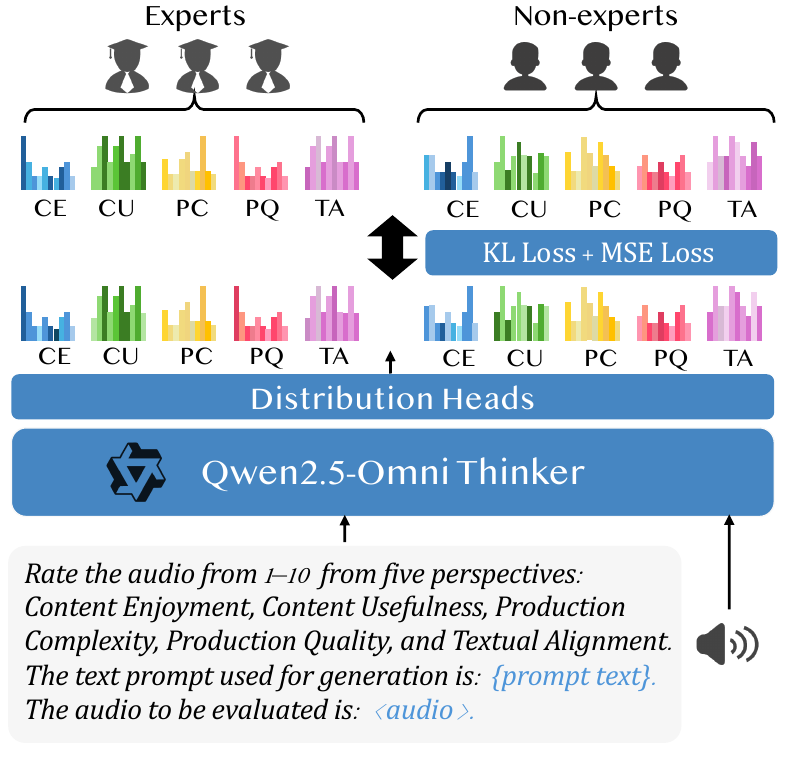}}
  \caption{Overview of Qwen-DisQA for TTA quality assessment, trained with distributional alignment.}
  \label{fig:overview}
\end{figure}

\section{Proposed Method}
\subsection{Problem Formulation}
On the top of AudioEval, We formulate TTA quality assessment as a multi-dimensional distribution prediction task. Given a text prompt $x^{(t)}$ and generated audio $x^{(a)}$, the goal is to predict perceptual ratings across five dimensions $\{d_1,\dots,d_5\}$ from two perspectives $v \in \{\text{expert}, \text{non-expert}\}$.  

For each $(d,v)$ pair, the target is a rating distribution $P_{d,v}(s)$ over scores $s \in \{1,\dots,10\}$. 
The model learns
\begin{equation}
    f(x^{(t)}, x^{(a)}) \;\rightarrow\; \{\hat{P}_{d,v}\}_{d,v},
\end{equation}
where $\hat{P}_{d,v}$ is the predicted distribution.  
Unlike traditional MOS regression that outputs a single scalar, our formulation preserves inter-rater variability, providing a richer and more reliable characterization of perceptual quality.

\subsection{Model Overview}

We propose \textbf{Qwen-DisQA}, a multimodal model for automatic TTA quality assessment. As depicted in Figure~\ref{fig:overview}, the model is built on Qwen2.5-Omni and takes as input both the text prompt $x^{(t)}$ and the generated audio $x^{(a)}$. We design a prompt template that explicitly integrates textual and acoustic information into a unified input sequence as shown in Figure~\ref{fig:overview}. The fused representation is then fed into task-specific prediction heads. Concretely, Qwen-DisQA employs ten independent heads, each corresponding to one dimension-perspective pair $(d,v)$. 
Each head is implemented as a linear projection layer followed by a softmax function, producing a probability distribution $\hat{P}_{d,v}(s)$ over discrete scores $s \in \{1,\dots,10\}$.

\subsection{Target Distribution}

For each $(d,v)$, three annotators provide discrete scores 
$y^{(m)} \in \{1,\dots,10\}$ $(m=1,2,3)$.  
Each score is mapped into a soft distribution over $k=1,\dots,10$ using a Gaussian kernel $p^{(m)}(k) \propto \exp\!\big(-\tfrac{1}{2}(\tfrac{y^{(m)}-k}{\sigma})^{2}\big)$. This soft labeling not only preserves individual annotations, but also captures the inherent uncertainty and semantic proximity between adjacent scores. 

The final target distribution is obtained by averaging across annotators:
\begin{equation}
\vspace{-4pt}
    P_{d,v}(k) = \frac{1}{3} \sum_{m=1}^{3} p^{(m)}(k), 
    \quad k=1,\dots,10.
\vspace{-4pt}
\end{equation}

\subsection{Training Targets}

Our loss combines distribution matching and mean regression.  
For each dimension--perspective pair $(d,v)$, we minimize the KL divergence between predicted and empirical distributions, together with the mean squared error (MSE) between predicted and ground-truth average scores:
\begin{equation}
    \mathcal{L} = \sum_{d,v} \Big[ \alpha \cdot D_{\text{KL}}\!\left(P_{d,v} \,\|\, \hat{P}_{d,v}\right) 
    + \lambda \cdot \big(\mu_{d,v} - \hat{\mu}_{d,v}\big)^2 \Big],
\end{equation}
where $\mu_{d,v}$ and $\hat{\mu}_{d,v}$ denote the ground-truth and predicted mean scores, respectively, and $\alpha$ and $\lambda$ control the balance between the two terms. The KL divergence term encourages the model to capture the full distribution of human ratings, preserving inter-rater variability and reflecting subjective uncertainty. The mean squared error term enforces accurate prediction of the expected rating, ensuring alignment with central perceptual tendencies. Together, these objectives enable Qwen-DisQA to deliver both distribution-aware and mean-consistent quality assessments.

\begin{table}[t]
\caption{Data split statistics. Counts are reported for samples, duration, unique systems (Sys.), and unseen systems.}
\label{tab:data-split-details}
\vspace{-5pt}
\begin{center}
\begin{small}
\setlength{\tabcolsep}{3.5pt}
\begin{tabular}{lrrrr}
\toprule
Set & \#Samples & Duration(h) & \#Sys. & \#Unseen Sys.\\
\midrule
Train & 3360 & 9.42 & 13  & -- \\
Val   &  420 & 1.17 & 17  & 5  \\
Test  &  420 & 1.12 & 17  & 6  \\
\bottomrule
\end{tabular}
\end{small}
\end{center}
\vspace{-10pt}
\end{table}

\newcolumntype{Y}{>{\centering\arraybackslash}X}

\begin{table*}[t]
  \centering
  \small
  \caption{Utterance-level PCC results of different systems. Models marked with ``*'' denote direct evaluation without fine-tuning, ``${\dagger}$'' indicates fine-tuning on pretrained CLAP encoder, 
and ``${\ddagger}$'' corresponds to LoRA fine-tuning on LLM.}
  \label{tab:results}

  \begin{tabularx}{\textwidth}{l *{5}{Y}:*{5}{Y}}
    \toprule
    \multirow{2}{*}{\textbf{Model}}
      & \multicolumn{5}{c:}{\textbf{Expert}} & \multicolumn{5}{c}{\textbf{Non-Expert}} \\
    \cmidrule(lr){2-6}\cmidrule(l){7-11}
     & CE & CU & PC & PQ & TA & CE & CU & PC & PQ & TA \\
    \midrule
    CLAP $^{*}$       & \textemdash & \textemdash & \textemdash & \textemdash & 0.341 
               & \textemdash & \textemdash & \textemdash & \textemdash & 0.386 \\
    Audiobox-Aesthetics $^{*}$  & 0.531 & 0.217 & 0.538 & 0.280 & \textemdash 
               & 0.255 & 0.363 & 0.223 & 0.306 & \textemdash \\
    \hdashline
    MusicEval-baseline $^{\dagger}$ & 0.452 & 0.468 & 0.418 & 0.413 & 0.499 
               & 0.218 & 0.281 & 0.312 & 0.342 & 0.531 \\
    Audio-Clap-finetune $^{\dagger}$  & 0.543 & 0.557 & 0.479 & 0.539 & 0.561 
               & 0.316 & 0.405 & 0.372 & 0.465 & 0.578 \\
    \hdashline
    Qwen2.5-Omni +R $^{\ddagger}$  &0.704&0.744&0.687&0.700&0.678&\underline{0.656}&\underline{0.725} &0.622&\underline{0.729}&\underline{0.731} \\
    Qwen2.5-Omni +KL $^{\ddagger}$    & \underline{0.718} & \textbf{0.752}& \underline{0.718} & \underline{0.712} & \textbf{0.731} & 0.639& \underline{0.725}& \textbf{0.652}& 0.708& 0.719 \\
    Qwen-DisQA $^{\ddagger}$ & \textbf{0.725} & \textbf{0.752} & \textbf{0.724} & \textbf{0.726} & \underline{0.704} & \textbf{0.671} & \textbf{0.735} & \textbf{0.652} & \textbf{0.738} & \textbf{0.742}\\
    \bottomrule
  \end{tabularx}
\end{table*}

\newcolumntype{Y}{>{\centering\arraybackslash}X}

\begin{table*}[t]
  \centering
  \small
  \caption{System-level PCC results of different systems. Models marked with ``*'' denote direct evaluation without fine-tuning, ``${\dagger}$'' indicates fine-tuning on pretrained CLAP encoder, and ``${\ddagger}$'' corresponds to LoRA fine-tuning on LLM.}
  \label{tab:systempcc}

  \begin{tabularx}{\textwidth}{l *{5}{Y}:*{5}{Y}}
    \toprule
    \multirow{2}{*}{\textbf{Model}}
      & \multicolumn{5}{c:}{\textbf{Expert}} & \multicolumn{5}{c}{\textbf{Non-Expert}} \\
    \cmidrule(lr){2-6}\cmidrule(l){7-11}
     & CE & CU & PC & PQ & TA & CE & CU & PC & PQ & TA \\
    \midrule
    CLAP $^{*}$       & \textemdash & \textemdash & \textemdash & \textemdash & 0.748 
               & \textemdash & \textemdash & \textemdash & \textemdash & 0.734 \\
    Audiobox-Aesthetics $^{*}$  & \textbf{0.813} & 0.737 & 0.702 & 0.706 & \textemdash 
               & 0.648 & 0.738 & 0.418 & 0.710 & \textemdash \\
    \hdashline
    MusicEval-baseline $^{\dagger}$ & 0.385 & 0.419 & 0.370 & 0.471 & 0.611 
               & 0.270 & 0.309 & 0.566 & 0.402 & 0.736 \\
    Audio-Clap-finetune $^{\dagger}$  & 0.446 & 0.444 & 0.367 & 0.633 & 0.705 
               & 0.342 & 0.406 & 0.620 & 0.470 & 0.791 \\
    \hdashline
    Qwen2.5-Omni +R $^{\ddagger}$  &0.692&0.803&0.616&0.798&0.817&\textbf{0.845}&\textbf{0.898}&\underline{0.811}&\textbf{0.925}&\underline{0.914} \\
    Qwen2.5-Omni +KL $^{\ddagger}$    & \underline{0.780} & \textbf{0.853}& \textbf{0.765}& \underline{0.836}& \textbf{0.900} & 0.678& 0.812& 0.765& 0.822& 0.872 \\
    Qwen-DisQA $^{\ddagger}$ & 0.777 & \underline{0.810} & \underline{0.726} & \textbf{0.848} & \underline{0.887} & \underline{0.799} & \underline{0.861} & \textbf{0.834} & \underline{0.876} & \textbf{0.920}\\
    \bottomrule
  \end{tabularx}
\end{table*}

\section{Experiments}

\subsection{Experimental Details}
% This section outlines the dataset partitioning, training setup, and evaluation metrics employed in our experiments.

\paragraph{Dataset Split.} 
We split the AudioEval dataset into training, validation, and test sets. Table~\ref{tab:data-split-details} summarizes the statistics for each split. The training set contains 3,360 samples from 13 systems, totaling 9.42 hours of audio. The validation and test sets each contain 420 samples, with durations of approximately 1.1 hours each. Both include 17 systems and over 250 prompts, among which 5 and 6 systems are unseen in the training set, respectively. This split design supports cross-system evaluation, enabling robust testing of model performance in unseen conditions.

\paragraph{Training Configuration.} 
We fine-tune our model, Qwen-DisQA, on the Qwen2.5-Omni-3B, backbone using parameter-efficient tuning. Specifically, we apply LoRA with a rank of 8 and train for 10 epochs using a batch size of 64 and a learning rate of 5e-4. The training objective combines KL divergence and MSE losses, weighted at 0.8 and 1.0 respectively. For the soft label distribution constructed, we set the standard deviation to $\sigma=0.15$. We monitor validation loss during training and select the checkpoint with the lowest score for final evaluation.

\paragraph{Evaluation Metrics.} We evaluate model predictions at two levels of granularity: the utterance level, which compares predicted and ground-truth scores for individual audio samples, and the system level, which aggregates scores across all outputs from each TTA system. We report two complementary metrics:
(1) Mean Squared Error (MSE): This measures the average squared difference between predicted scores and human annotations. A lower MSE indicates better numerical accuracy in score prediction.
(2) Pearson Correlation Coefficient (PCC): This quantifies the linear correlation between predicted and true scores, capturing ranking consistency. A higher PCC reflects better alignment with human judgment.

\subsection{Compared Approaches}
We evaluated three types of models on the AudioEval.

\textbf{Pre-trained evaluators.}
We first evaluate the zero-shot performance of two widely used models:  Audiobox-Aesthetics~\cite{aes} and CLAP\footnote{\url{https://huggingface.co/microsoft/msclap/blob/main/CLAP\_weights\_2023.pth}}. Both have been adopted in prior work as automatic metrics for Text-to-Audio generation. We test them directly to assess whether their outputs align with human across multiple dimensions.

\textbf{CLAP-based fine-tuned models.}
We implement two supervised baselines by adding task-specific heads to the CLAP encoder and training them on the AudioEval dataset: MusicEval-baseline\footnote{ \url{https://huggingface.co/lukewys/laion\_clap/blob/main/music\_audioset\_epoch\_15\_esc\_90.14.pt}} and Audio-Clap-finetune\footnote{ \url{https://huggingface.co/lukewys/laion\_clap/blob/main/630k-best.pt}}. These models represent typical regression-based approaches adapted from contrastive pre-training.

\textbf{Large-language-model evaluators.}
Our method, Qwen-DisQA, is based on a large multimodal language model trained with both KL divergence and MSE losses to capture score distributions. To analyze the impact of each component, we include two ablation variants that use only KL (+KL) or only MSE (+R) loss during training.

\begin{figure*}[t]
  \centering
  \includegraphics[width=\linewidth]{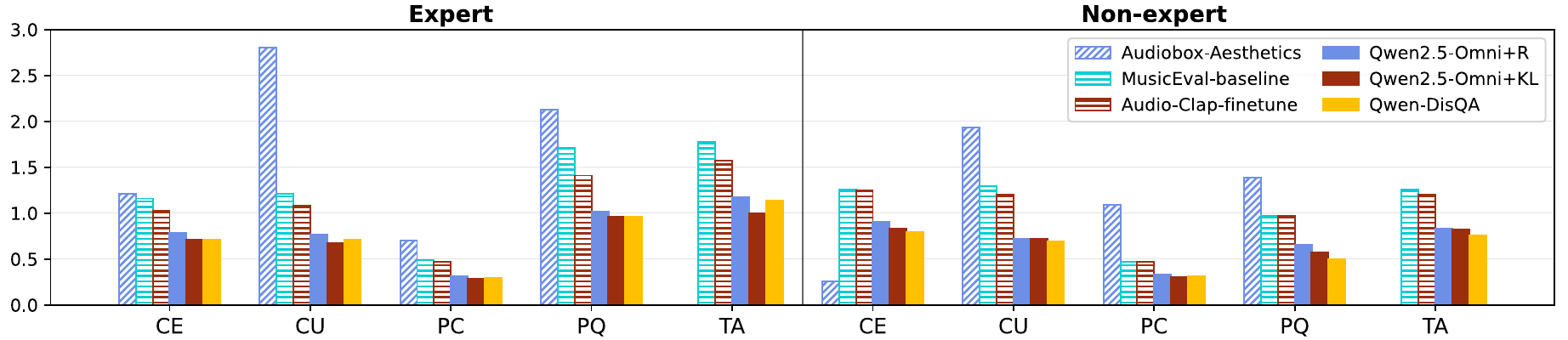}
  \vspace{-8pt}
  \caption{Utterance-level MSE results of different systems. Different bar hatch patterns are used to distinguish system types.}
  \label{fig:uttmse}
  \vspace{-6pt}
\end{figure*}

\subsection{Correlation Analysis}

Table~\ref{tab:results} reports utterance-level PCC between model predictions and human scores across five evaluation dimensions and two rater views. Qwen-DisQA consistently outperforms all baselines, achieving the highest or near-highest PCCs in most dimensions under both expert and non-expert perspectives. It reaches average PCCs of 0.726 and 0.708 for the expert and non-expert views, respectively, confirming its effectiveness in modeling fine-grained, multi-dimensional human judgments. Ablation models with only KL or only MSE loss also perform competitively, suggesting both objectives contribute meaningfully. Models without fine-tuning, such as CLAP and Audiobox-Aesthetics, show substantially lower correlations, highlighting the importance of adapting evaluators to the AudioEval dataset. Additionally, expert-view scores are generally easier to predict, likely due to their lower variance and more structured evaluation behavior.

Table~\ref{tab:systempcc} presents the Pearson correlation coefficients between predicted and human scores at the system level, averaged across all utterances per system. Qwen-DisQA achieves the strongest overall performance, with average PCCs of 0.848 (expert) and 0.862 (non-expert), outperforming all baseline models across most evaluation dimensions. Particularly under the non-expert view, it reaches 0.920 correlation in Textual Alignment and over 0.8 in all other dimensions, showing its robustness in capturing holistic system behavior. In contrast, zero-shot models such as CLAP and Audiobox-Aesthetics yield competitive scores on specific dimensions (e.g., TA), but suffer from inconsistent performance elsewhere, revealing their limited generalization without fine-tuning. Notably, CLAP performs surprisingly well in Textual Alignment (0.748 expert), but underperforms in other aspects like Content Usefulness. Fine-tuned CLAP-based models show slight improvements but still lag behind large language model-based methods. Compared to its KL-only and MSE-only variants, Qwen-DisQA consistently yields higher correlations, confirming the benefit of modeling distributional supervision via a combined objective.

\subsection{MSE Analysis}

Figure~\ref{fig:uttmse} presents the utterance-level mean squared error of different models across five evaluation dimensions, from both expert and non-expert perspectives. Among all systems, Qwen-DisQA consistently achieves the lowest MSE in most dimensions and under both views, indicating its strong ability to approximate fine-grained human judgments. Notably, zero-shot models such as Audiobox-Aesthetics and MusicEval-baseline suffer from high errors, especially in Production Quality and Content Usefulness, suggesting limited generalizability without adaptation. Fine-tuned CLAP variants reduce error moderately, but still lag behind large language model-based systems. The Qwen2.5-based methods show strong performance, and Qwen-DisQA further improves over its ablated variants by jointly optimizing KL and MSE losses. These findings demonstrate that Qwen-DisQA not only achieves high correlation with human ratings but also produces numerically accurate predictions.

\begin{figure}[tb]
  \centering
  \centerline{\includegraphics[width=1\linewidth]{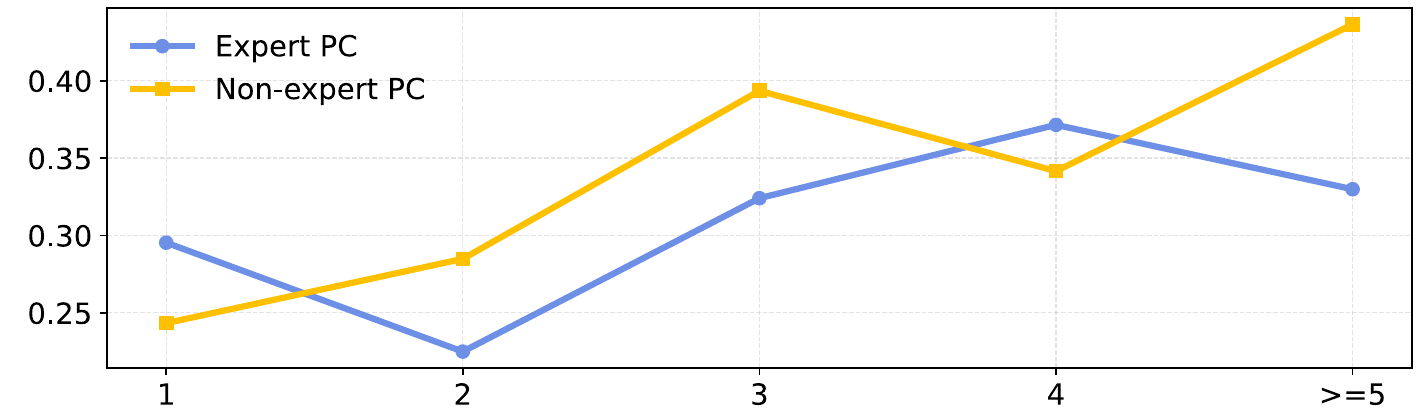}}
  \caption{MSE of Production Complexity prediction for expert vs. non-expert across prompts with different numbers of sound events.}
  \label{fig:PCdim_analysis}
  \vspace{-4pt}
\end{figure}

\begin{figure}[tb]
  \centering
  \centerline{\includegraphics[width=1\linewidth]{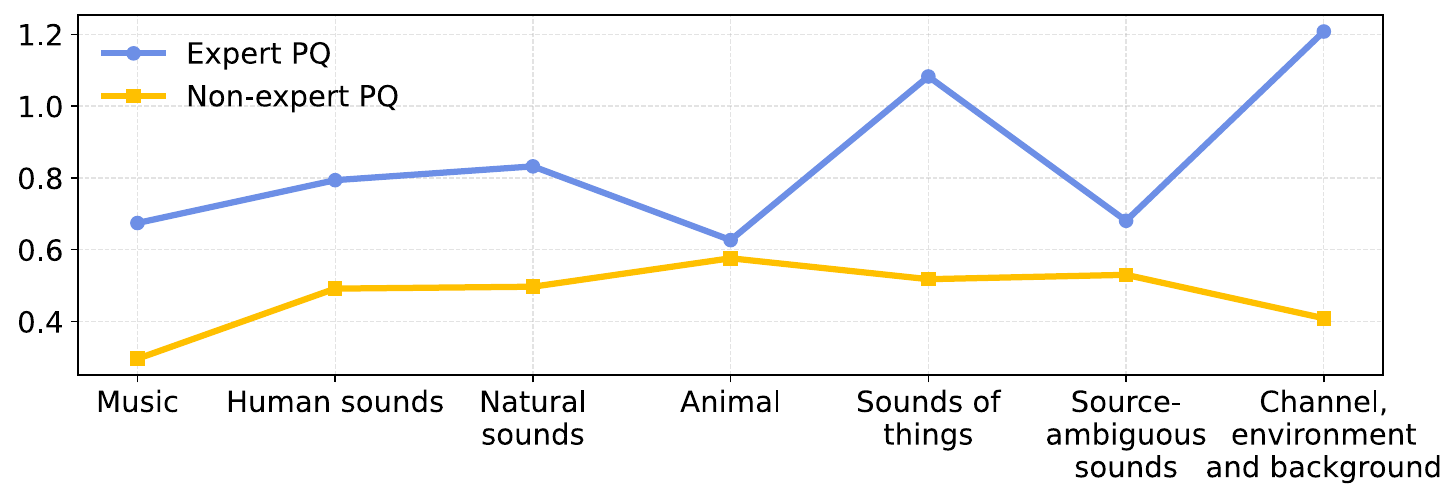}}
  \caption{MSE of production quality prediction for expert vs. non-expert ratings across sound-type categories.}
  \vspace{-8pt}
  \label{fig:PQdim_analysis}

\end{figure}

\subsection{Error Analysis}

As shown in Figure~\ref{fig:PCdim_analysis}, both expert and non-expert MSE tend to increase as the number of sound events grows from 1 to 4, indicating a shared pattern: denser, more multi-event prompts are intrinsically harder to model for Production Complexity, likely because they introduce more overlapping cues (layering, simultaneity, masking) and greater variability in how “complexity” is perceived.

As shown in Figure~\ref{fig:PQdim_analysis}, non-expert PQ MSE is consistently lower and relatively stable across categories, whereas expert PQ MSE is higher and varies substantially by sound type, with clear error peaks for Sounds of things and Channel/environment/background. This pattern is consistent with non-experts primarily using coarse, broadly applicable cues (e.g., overall clarity, obvious noise/distortion), which generalize similarly across content. Experts, however, tend to incorporate finer and more content-conditional factors. These additional, category-dependent criteria increase the effective complexity of the target and make expert PQ harder to predict, particularly for heterogeneous or acoustically complex sound types.

\section{Conclusion}

In this work, we introduced AudioEval, the first large-scale multi-dimensional dataset for text-to-audio evaluation, annotated by both experts and non-experts across five perceptual dimensions. Building upon this resource, we proposed Qwen-DisQA, a multimodal scoring model that predicts human-like quality ratings from text--audio pairs. Experimental results show that different types of automatic evaluators achieve varying levels of correlation and robustness on AudioEval, and that Qwen-DisQA provides a strong reference baseline under our evaluation setting. Finally, leveraging AudioEval, we also conduct error analysis across perceptual dimensions and sound types, offering insights into failure modes and directions for future improvement.

\clearpage

\section*{Impact Statement}

This paper presents work whose goal is to advance the field of Machine
Learning. There are many potential societal consequences of our work, none
which we feel must be specifically highlighted here.

\nocite{langley00}

\bibliography{example_paper}

@inproceedings{langley00,
 author    = {P. Langley},
 title     = {Crafting Papers on Machine Learning},
 year      = {2000},
 pages     = {1207--1216},
 editor    = {Pat Langley},
 booktitle     = {Proceedings of the 17th International Conference
              on Machine Learning (ICML 2000)},
 address   = {Stanford, CA},
 publisher = {Morgan Kaufmann}
}

@misc{huang2023make2,
      title={Make-An-Audio 2: Temporal-Enhanced Text-to-Audio Generation}, 
      author={Jiawei Huang and Yi Ren and Rongjie Huang and Dongchao Yang and Zhenhui Ye and Chen Zhang and Jinglin Liu and Xiang Yin and Zejun Ma and Zhou Zhao},
      year={2023},
      eprint={2305.18474},
      archivePrefix={arXiv},
      primaryClass={cs.SD},
      url={https://arxiv.org/abs/2305.18474}, 
}

@misc{huang2023make1,
      title={Make-An-Audio: Text-To-Audio Generation with Prompt-Enhanced Diffusion Models}, 
      author={Rongjie Huang and Jiawei Huang and Dongchao Yang and Yi Ren and Luping Liu and Mingze Li and Zhenhui Ye and Jinglin Liu and Xiang Yin and Zhou Zhao},
      year={2023},
      eprint={2301.12661},
      archivePrefix={arXiv},
      primaryClass={cs.SD},
      url={https://arxiv.org/abs/2301.12661}, 
}

@article{auffusion,
  author  = {Xue, Jinlong and Deng, Yayue and Gao, Yingming and Li, Ya},
  title   = {Auffusion: Leveraging the Power of Diffusion and Large Language Models for Text-to-Audio Generation},
  journal = {IEEE/ACM Trans. Audio Speech Lang. Process.},
  year    = {2024},
  volume  = {32},
  pages   = {4700--4712}
}

@inproceedings{tango,
  title={Text-to-audio generation using instruction guided latent diffusion model},
  author={Ghosal, Deepanway and Majumder, Navonil and Mehrish, Ambuj and Poria, Soujanya},
  booktitle={Proceedings of the 31st ACM International Conference on Multimedia},
  pages={3590--3598},
  year={2023}
}

@inproceedings{tango2,
  title={Tango 2: Aligning diffusion-based text-to-audio generations through direct preference optimization},
  author={Majumder, Navonil and Hung, Chia-Yu and Ghosal, Deepanway and Hsu, Wei-Ning and Mihalcea, Rada and Poria, Soujanya},
  booktitle={Proceedings of the 32nd ACM International Conference on Multimedia},
  pages={564--572},
  year={2024}
}

@inproceedings{audiogen,
  title = {AudioGen: Textually Guided Audio Generation},
  author = {Felix Kreuk and Gabriel Synnaeve and Adam Polyak and Uriel Singer and Alexandre D{\'e}fossez and Jade Copet and Devi Parikh and Yaniv Taigman and Yossi Adi},
  year = {2023},
  booktitle = {Proc. ICLR},
}

@inproceedings{liu2023audioldm,
  title     = {AudioLDM: Text-to-Audio Generation with Latent Diffusion Models},
  author    = {Liu, Haohe and Chen, Zehua and Yuan, Yi and Mei, Xinhao and Liu, Xubo and Mandic, Danilo and Wang, Wenwu and Plumbley, Mark D},
  booktitle = {Proc. ICML},
  pages     = {21450--21474},
  year      = {2023}
}

@article{audioldm2-2024taslp,
  title={Audioldm 2: Learning holistic audio generation with self-supervised pretraining},
  author={Liu, Haohe and Yuan, Yi and Liu, Xubo and Mei, Xinhao and Kong, Qiuqiang and Tian, Qiao and Wang, Yuping and Wang, Wenwu and Wang, Yuxuan and Plumbley, Mark D},
  journal={IEEE/ACM Transactions on Audio, Speech, and Language Processing},
  volume={32},
  pages={2871--2883},
  year={2024},
  publisher={IEEE}
}

@misc{ziv2024magnet,
      title={Masked Audio Generation using a Single Non-Autoregressive Transformer}, 
      author={Alon Ziv and Itai Gat and Gael Le Lan and Tal Remez and Felix Kreuk and Alexandre Défossez and Jade Copet and Gabriel Synnaeve and Yossi Adi},
      year={2024},
      eprint={2401.04577},
      archivePrefix={arXiv},
      primaryClass={cs.SD},
      url={https://arxiv.org/abs/2401.04577}, 
}

@inproceedings{evans2024stableaudio,
author = {Evans, Zach and Carr, CJ and Taylor, Josiah and Hawley, Scott H. and Pons, Jordi},
title = {Fast timing-conditioned latent audio diffusion},
year = {2024},
publisher = {JMLR.org},
booktitle = {Proceedings of the 41st International Conference on Machine Learning},
articleno = {505},
numpages = {14},
location = {Vienna, Austria},
series = {ICML'24}
}

@misc{CLAP2023,
      title={Natural Language Supervision for General-Purpose Audio Representations}, 
      author={Benjamin Elizalde and Soham Deshmukh and Huaming Wang},
      year={2023},
      eprint={2309.05767},
      archivePrefix={arXiv},
      primaryClass={cs.SD},
      url={https://arxiv.org/abs/2309.05767}
}

@article{aes,
  title={Meta audiobox aesthetics: Unified automatic quality assessment for speech, music, and sound},
  author={Tjandra, Andros and Wu, Yi-Chiao and Guo, Baishan and Hoffman, John and Ellis, Brian and Vyas, Apoorv and Shi, Bowen and Chen, Sanyuan and Le, Matt and Zacharov, Nick and others},
  journal={arXiv preprint arXiv:2502.05139},
  year={2025}
}

@inproceedings{audioset,
  title={Audio set: An ontology and human-labeled dataset for audio events},
  author={Gemmeke, Jort F and Ellis, Daniel PW and Freedman, Dylan and Jansen, Aren and Lawrence, Wade and Moore, R Channing and Plakal, Manoj and Ritter, Marvin},
  booktitle={2017 IEEE international conference on acoustics, speech and signal processing (ICASSP)},
  pages={776--780},
  year={2017},
  organization={IEEE}
}

@misc{fad,
      title={Fr\'echet Audio Distance: A Metric for Evaluating Music Enhancement Algorithms}, 
      author={Kevin Kilgour and Mauricio Zuluaga and Dominik Roblek and Matthew Sharifi},
      year={2019},
      eprint={1812.08466},
      archivePrefix={arXiv},
      primaryClass={eess.AS},
      url={https://arxiv.org/abs/1812.08466}, 
}

@inproceedings{wang23r_interspeech,
  title     = {RAMP: Retrieval-Augmented MOS Prediction via Confidence-based Dynamic Weighting},
  author    = {Wang, Hui and Zhao, Shiwan and Zheng, Xiguang and Qin, Yong},
  year      = {2023},
  booktitle = {INTERSPEECH 2023},
  pages     = {1095--1099},
  doi       = {10.21437/Interspeech.2023-851},
  issn      = {2958-1796},
}

@inproceedings{lafma,
  title     = {LAFMA: A Latent Flow Matching Model for Text-to-Audio Generation},
  author    = {Guan, Wenhao and Wang, Kaidi and Zhou, Wangjin and Wang, Yang and Deng, Feng and Wang, Hui and Li, Lin and Hong, Qingyang and Qin, Yong},
  year      = {2024},
  booktitle = {Interspeech 2024},
  pages     = {4813--4817},
  doi       = {10.21437/Interspeech.2024-1848},
  issn      = {2958-1796},
}

@INPROCEEDINGS{musiceval,
  author={Liu, Cheng and Wang, Hui and Zhao, Jinghua and Zhao, Shiwan and Bu, Hui and Xu, Xin and Zhou, Jiaming and Sun, Haoqin and Qin, Yong},
  booktitle={Proc. ICASSP}, 
  title={MusicEval: A Generative Music Dataset with Expert Ratings for Automatic Text-to-Music Evaluation}, 
  year={2025},
  volume={},
  number={},
  pages={1-5},}

@article{valle,
  title={Neural codec language models are zero-shot text to speech synthesizers},
  author={Chen, Sanyuan and Wang, Chengyi and Wu, Yu and Zhang, Ziqiang and Zhou, Long and Liu, Shujie and et al., Huaming and Li, Jinyu and others},
  journal={IEEE Transactions on Audio, Speech and Language Processing},
  year={2025},
  publisher={IEEE}
}

@inproceedings{fid,
 author = {Heusel, Martin and Ramsauer, Hubert and Unterthiner, Thomas and Nessler, Bernhard and Hochreiter, Sepp},
 booktitle = {Proc. NeurIPS},
 pages = {},
 title = {GANs Trained by a Two Time-Scale Update Rule Converge to a Local Nash Equilibrium},
 volume = {30},
 year = {2017}
}

@inproceedings{fad-correlates-poorly,
  title={Evaluating Generative Audio Systems and Their Metrics},
  booktitle={{Proc. ISMIR}},
  author={Vinay, Ashvala and Lerch, Alexander},
  year={2022}
}

@inproceedings{utmos,
  title     = {UTMOS: UTokyo-SaruLab System for VoiceMOS Challenge 2022},
  author    = {Saeki, Takaaki and Xin, Detai and Nakata, Wataru and Koriyama, Tomoki and Takamichi, Shinnosuke and Saruwatari, Hiroshi},
  booktitle = {Proc. Interspeech},
  pages     = {4521--4525},
  year      = {2022}
}

@article{songeval,
  title={SongEval: A Benchmark Dataset for Song Aesthetics Evaluation},
  author={Yao, Jixun and Ma, Guobin and Xue, Huixin and Chen, Huakang and Hao, Chunbo and Jiang, Yuepeng and Liu, Haohe and Yuan, Ruibin and Xu, Jin and Xue, Wei and others},
  journal={arXiv preprint arXiv:2505.10793},
  year={2025}
}

@article{qwen25omni,
  title={Qwen2. 5-omni technical report},
  author={Xu, Jin and Guo, Zhifang and He, Jinzheng and Hu, Hangrui and He, Ting and Bai, Shuai and Chen, Keqin and Wang, Jialin and Fan, Yang and Dang, Kai and others},
  journal={arXiv preprint arXiv:2503.20215},
  year={2025}
}

@article{consistencytta,
  title={Consistencytta: Accelerating diffusion-based text-to-audio generation with consistency distillation},
  author={Bai, Yatong and Dang, Trung and Tran, Dung and Koishida, Kazuhito and Sojoudi, Somayeh},
  journal={arXiv preprint arXiv:2309.10740},
  year={2023}
}

@inproceedings{audiolcm,
  title={Audiolcm: Efficient and high-quality text-to-audio generation with minimal inference steps},
  author={Liu, Huadai and Huang, Rongjie and Liu, Yang and Cao, Hengyuan and Wang, Jialei and Cheng, Xize and Zheng, Siqi and Zhao, Zhou},
  booktitle={Proc. ACM MM},
  pages={7008--7017},
  year={2024}
}

@inproceedings{ctag,
  author    = {Cherep, Manuel and Singh, Nikhil and Shand, Jessica},
  title     = {Creative text-to-audio generation via synthesizer programming},
  booktitle = {Proc. ICML},
  year      = {2024}
}

@inproceedings{picoaudio,
  title     = {PicoAudio: Enabling Precise Temporal Controllability in Text-to-Audio Generation},
  author    = {Xie, Zeyu and Xu, Xuenan and Wu, Zhizheng and Wu, Mengyue},
  booktitle = {Proc. ICASSP},
  pages     = {1--5},
  year      = {2025}
}

@inproceedings{ezaudio,
  title     = {{EzAudio: Enhancing Text-to-Audio Generation with Efficient Diffusion Transformer}},
  author    = {Jiarui Hai and Yong Xu and Hao Zhang and Chenxing Li and Helin Wang and Mounya Elhilali and Dong Yu},
  year      = {2025},
  booktitle = {{Interspeech 2025}},
  pages     = {4233--4237},
  doi       = {10.21437/Interspeech.2025-1137},
  issn      = {2958-1796},
}

@inproceedings{soundctm,
  title     = {SoundCTM: Uniting Score-based and Consistency Models for Text-to-Sound Generation},
  author    = {Saito, Koichi and Kim, Dongjun and Shibuya, Takashi and Lai, Chieh-Hsin and Zhong, Zhi and Takida, Yuhta and Mitsufuji, Yuki},
  booktitle = {Proc. NeurIPS Workshop},
  year      = {2024}
}

@inproceedings{luminat2x,
  title     = {Lumina-T2X: Scalable Flow-based Large Diffusion Transformer for Flexible Resolution Generation},
  author    = {Gao, Peng and Zhuo, Le and Liu, Dongyang and Du, Ruoyi and Luo, Xu and Qiu, Longtian and Zhang, Yuhang and Huang, Rongjie and Geng, Shijie and Zhang, Renrui and others},
  booktitle = {Proc. ICLR},
  year      = {2025},
  url       = {https://openreview.net/forum?id=EbWf36quzd}
}

@inproceedings{flashaudio,
  title     = {FlashAudio: Rectified Flow for Fast and High-Fidelity Text-to-Audio Generation},
  author    = {Liu, Huadai and Wang, Jialei and Huang, Rongjie and Liu, Yang and Lu, Heng and Zhao, Zhou and Xue, Wei},
  booktitle = {Proc. ACL},
  pages     = {13694--13710},
  year      = {2025}
}

@inproceedings{t2afeedback,
  title     = {T2A-Feedback: Improving Basic Capabilities of Text-to-Audio Generation via Fine-grained AI Feedback},
  author    = {Wang, Zehan and Lei, Ke and Zhu, Chen and Huang, Jiawei and Zhou, Sashuai and Liu, Luping and Cheng, Xize and Ji, Shengpeng and Ye, Zhenhui and Jin, Tao and others},
  booktitle = {Proc. ACL},
  pages     = {23535--23547},
  year      = {2025}
}

@article{bvcc,
  title={How do voices from past speech synthesis challenges compare today?},
  author={Cooper, Erica and Yamagishi, Junichi},
  journal={arXiv preprint arXiv:2105.02373},
  year={2021}
}

@article{audiox,
  title={AudioX: Diffusion Transformer for Anything-to-Audio Generation},
  author={Tian, Zeyue and Jin, Yizhu and Liu, Zhaoyang and Yuan, Ruibin and Tan, Xu and Chen, Qifeng and Xue, Wei and Guo, Yike},
  journal={arXiv preprint arXiv:2503.10522},
  year={2025}
}

@article{ttabench,
  title={TTA-Bench: A Comprehensive Benchmark for Evaluating Text-to-Audio Models},
  author={Wang, Hui and Liu, Cheng and Chen, Junyang and Liu, Haoze and Jia, Yuhang and Zhao, Shiwan and Zhou, Jiaming and Sun, Haoqin and Bu, Hui and Qin, Yong},
  journal={arXiv preprint arXiv:2509.02398},
  year={2025}
}

@INPROCEEDINGS{ARC-TTA,
  author={Novack, Zachary and Evans, Zach and Zukowski, Zack and Taylor, Josiah and Carr, CJ and Parker, Julian and Al-Sinan, Adnan and Iodice, Gian Marco and McAuley, Julian and Berg-Kirkpatrick, Taylor and Pons, Jordi},
  booktitle={2025 IEEE Workshop on Applications of Signal Processing to Audio and Acoustics (WASPAA)}, 
  title={Fast Text-to-Audio Generation with Adversarial Post-Training}, 
  year={2025},
  volume={},
  number={},
  pages={1-5},
  doi={10.1109/WASPAA66052.2025.11230941}}

@inproceedings{shi2025audiocache,
  title={Audiocache: Accelerate audio generation with training-free layer caching},
  author={Shi, Qingyang and Du, Zhicheng and Lu, Jiasheng and Liang, Yingshan and Zhang, Xinyu and Wang, Yiran and Peng, Jing and Yuan, Kehong},
  booktitle={ICASSP 2025-2025 IEEE International Conference on Acoustics, Speech and Signal Processing (ICASSP)},
  pages={1--5},
  year={2025},
  organization={IEEE}
}

@inproceedings{InfiniteAudio,
  title     = {{InfiniteAudio: Infinite-Length Audio Generation with Consistency}},
  author    = {Chaeyoung Jung and Hojoon Ki and Ji-Hoon Kim and Junmo Kim and Joon Son Chung},
  year      = {2025},
  booktitle = {{Interspeech 2025}},
  pages     = {4213--4217},
  doi       = {10.21437/Interspeech.2025-209}  ,
  issn      = {2958-1796},
}

@inproceedings{felle,
author = {Wang, Hui and Liu, Shujie and Meng, Lingwei and Li, Jinyu and Yang, Yifan and Zhao, Shiwan and Sun, Haiyang and Liu, Yanqing and Sun, Haoqin and Zhou, Jiaming and Lu, Yan and Qin, Yong},
title = {FELLE: Autoregressive Speech Synthesis with Token-Wise Coarse-to-Fine Flow Matching},
year = {2025},
isbn = {9798400720352},
publisher = {Association for Computing Machinery},
address = {New York, NY, USA},
url = {https://doi.org/10.1145/3746027.3755494},
doi = {10.1145/3746027.3755494},
booktitle = {Proceedings of the 33rd ACM International Conference on Multimedia},
pages = {10229–10238},
numpages = {10},
location = {Dublin, Ireland},
series = {MM '25}
}

@inproceedings{chiang23_interspeech,
  title     = {Why We Should Report the Details in Subjective Evaluation of TTS More Rigorously},
  author    = {Cheng-Han Chiang and Wei-Ping Huang and Hung-yi Lee},
  year      = {2023},
  booktitle = {Interspeech 2023},
  pages     = {5551--5555},
  doi       = {10.21437/Interspeech.2023-416},
  issn      = {2958-1796},
}

@article{10.1145/3769106,
author = {Lerch, Alexander and Arthur, Claire and Bryan-Kinns, Nick and Ford, Corey and Sun, Qianyi and Vinay, Ashvala},
title = {Survey on the Evaluation of Generative Models in Music},
year = {2025},
issue_date = {March 2026},
publisher = {Association for Computing Machinery},
address = {New York, NY, USA},
volume = {58},
number = {4},
issn = {0360-0300},
url = {https://doi.org/10.1145/3769106},
doi = {10.1145/3769106},
journal = {ACM Comput. Surv.},
month = oct,
articleno = {99},
numpages = {36}
}
\bibliographystyle{icml2026}

%%%%%%%%%%%%%%%%%%%%%%%%%%%%%%%%%%%%%%%%%%%%%%%%%%%%%%%%%%%%%%%%%%%%%%%%%%%%%%%
%%%%%%%%%%%%%%%%%%%%%%%%%%%%%%%%%%%%%%%%%%%%%%%%%%%%%%%%%%%%%%%%%%%%%%%%%%%%%%%
% APPENDIX
%%%%%%%%%%%%%%%%%%%%%%%%%%%%%%%%%%%%%%%%%%%%%%%%%%%%%%%%%%%%%%%%%%%%%%%%%%%%%%%
%%%%%%%%%%%%%%%%%%%%%%%%%%%%%%%%%%%%%%%%%%%%%%%%%%%%%%%%%%%%%%%%%%%%%%%%%%%%%%%
\newpage
\appendix
\onecolumn
\section{AudioEval Dataset.}
\subsection{System Detail}
\label{app:system_list}
% 系统信息：绘制表格，id,模型名字（引用），年份，参数量，再加简单的介绍段落
Table~\ref{tab:system_overview} provides an overview of the text-to-audio generation systems evaluated in this work. The table summarizes representative autoregressive, diffusion-based, flow-based, and signal-processing-based models proposed between 2022 and 2025, along with their parameter scales and architectural types. This diverse system set allows for a comprehensive comparison across different modeling paradigms and design choices in modern text-to-audio generation.

\begin{table}[htbp]
\centering
\caption{Overview of Text-to-Audio Generation Systems}
\label{tab:system_overview}
\begin{tabular}{llm{2.2cm}ll}
\toprule
ID & TTA Model & Parameter Size & Year & Model Type \\
\midrule
S001 & AudioGen-medium \cite{audiogen} & 1.5B & 2022 & Autoregressive (AR) \\
S002 & AudioLDM-l-full \cite{liu2023audioldm} & 739M & 2023 & Latent Diffusion Model (LDM) \\
S003 & AudioLDM 2-large \cite{audioldm2-2024taslp} & 712M & 2023 & Latent Diffusion Model (LDM) \\
S004 & Auffusion-full \cite{auffusion} & 1.1B & 2024 & Latent Diffusion Model (LDM) \\
S005 & MAGNeT-medium \cite{ziv2024magnet} & 1.5B & 2024 & Non-Autoregressive (NAR) \\
S006 & Make-An-Audio \cite{huang2023make1} & 453M & 2023 & Latent Diffusion Model (LDM) \\
S007 & Make-An-Audio-2 \cite{huang2023make2} & 937M & 2023 & Latent Diffusion Model (LDM) \\
S008 & Stable Audio Open 1.0 \cite{evans2024stableaudio} & 1.21B & 2024 & Diffusion Transformer (DiT) \\
S009 & Tango-full \cite{tango} & 866M & 2023 & Latent Diffusion Model (LDM) \\
S010 & Tango-2-full \cite{tango2} & 866M & 2024 & Latent Diffusion Model (LDM) \\
S011 & AudioLCM \cite{audiolcm} & -- & 2024 & Latent Consistency Model (LCM) \\
S012 & EzAudio \cite{ezaudio} & 596M / 874M & 2024 & Diffusion Transformer (DiT) \\
S013 & PicoAudio \cite{picoaudio} & -- & 2024 & Conditional Diffusion Model (LDM) \\
S014 & AudioX \cite{audiox} & -- & 2025 & Diffusion Transformer (DiT) \\
S015 & ARC-TTA \cite{ARC-TTA} & 1.21B & 2025 & Diffusion Transformer (DiT) \\
S016 & T2A-Feedback \cite{t2afeedback} & 937M & 2025 & Latent Diffusion Model (LDM) \\
S017 & CTAG \cite{ctag} & -- & 2024 & Digital Signal Processing (DSP) \\
S018 & FlashAudio \cite{flashaudio} & 74M / 197M / 429M & 2024 & Rectified Flow \\
S019 & Lumina-T2X \cite{luminat2x} & 7B & 2024 & Flow-based DiT (Flag-DiT) \\
S020 & ConsistencyTTA \cite{consistencytta} & -- & 2023 & Latent Diffusion Model (LDM) \\
S021 & SoundCTM \cite{soundctm} & 1B & 2024 & Diffusion Transformer (DiT) \\
S022 & LAFMA \cite{lafma} & 272M & 2024 & Flow Matching LDM \\
S023 & AudioCache \cite{shi2025audiocache} & 937M & 2025 & Diffusion Transformer (DiT) \\
S024 & InfiniteAudio \cite{InfiniteAudio} & -- & 2025 & Latent Diffusion Model (LDM) \\
\bottomrule
\end{tabular}
\end{table}

% 布局1
\begin{wrapfigure}{r}{0.45\columnwidth}
    \vspace{-12mm}
    \includegraphics[width=0.45\columnwidth]{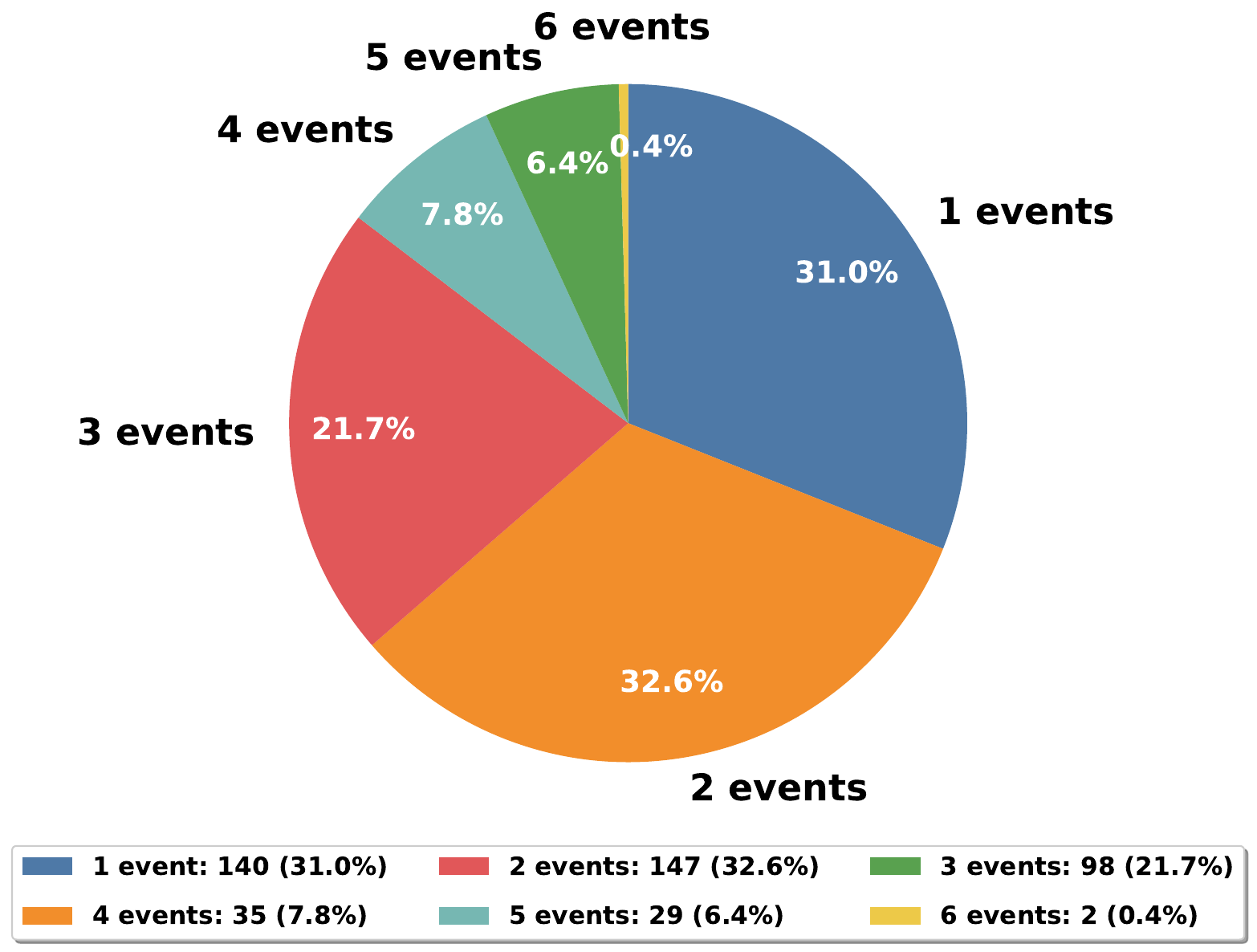}
    \vspace{-3mm}
    \caption{Distribution of Sound Event Count in Audio Prompts}
    \vspace{-4mm}
    \label{fig:placeholder}
\end{wrapfigure}

\subsection{Prompt Detail}
\label{app:prompt}
% new 再画个prompt示例表格，包括prompt内容，以及标记的几种属性
% prompt信息：绘制声音事件比例
Figure~\ref{fig:placeholder} illustrates the distribution of the number of sound events contained in the text prompts used in our dataset. Most prompts describe one or two sound events, accounting for approximately 64\% of all samples, while a substantial portion includes three events, reflecting moderate compositional complexity. Prompts with four or more sound events are less frequent, forming a long tail that captures more complex acoustic scenarios. This distribution ensures coverage of both simple and multi-event prompts, enabling evaluation of text-to-audio systems under varying levels of semantic and structural complexity. Annotated examples of text prompts with different numbers of sound events are provided in Table~\ref{tab:prompt_examples}, which summarizes representative prompts along with their associated scene categories, sound event counts, and hierarchical sound event labels.
\begin{table}[t]
\centering
\caption{Example prompts with scene categories and sound event annotations.}
\label{tab:prompt_examples}
\renewcommand{\arraystretch}{1.2}
% \begin{tabularx}{\linewidth}{c X X c X X}
\begin{tabularx}{\linewidth}{C{1.2cm}
p{3.4cm}
p{2.5cm}
C{2cm}
p{2.8cm}
p{3cm}}

\hline
\textbf{Prompt ID} &
\textbf{Prompt Text} &
\textbf{Scene Category} &
\textbf{Sound Event Count} &
\textbf{Level-1 Categories} &
\textbf{Level-2 Categories} \\
\hline
N001 &
A man speaks as birds chirp and dogs bark &
Daily life scenes &
3 &
Human sounds; Animal; Animal &
Human voice; Wild animals; Domestic animals and pets \\

N002 &
Thunder and a gentle rain &
Natural \& outdoor scenes &
2 &
Natural sounds; Natural sounds &
Thunderstorm; Water \\

N003 &
Birds are squawking, and ducks are quacking &
Natural \& outdoor scenes &
2 &
Animal; Animal &
Wild animals; Wild animals \\

N004 &
Repeated gunfire and screaming in the background &
Daily life scenes &
2 &
Sounds of things; Human sounds &
Explosion; Human voice \\

N005 &
An engine revving and then tires squealing &
Transportation \& travel scenes &
2 &
Sounds of things; Sounds of things &
Engine; Vehicle \\
\hline
\end{tabularx}
\end{table}

\subsection{Annotator Detail}
\label{app:rater}
% 标注人员信息：专家，非专家信息表格
The demographic characteristics of the non-expert and expert annotators involved in our study are summarized in Tables~\ref{tab:annotator_info} and~\ref{tab:expert_annotator_info}. The non-expert group consists of undergraduate-level participants from diverse academic backgrounds, including engineering, media, design, among others, with English proficiency ranging from undergraduate-level competence to CET-4 and CET-6 certifications, reflecting a realistic distribution of general listeners for the annotation tasks.

In contrast, the expert annotators are professionally trained individuals with formal education in music performance or music education, enabling more technically informed and consistent judgments of audio quality and content attributes. By including both non-expert and expert annotators, the study facilitates comparative analysis across different expertise levels and enhances the reliability and robustness of the evaluation results.

% 布局2
% \begin{wrapfigure}{r}{0.5\columnwidth}
%     \vspace{-14mm}
%     \includegraphics[width=0.5\columnwidth]{sound_event_distribution.pdf}
%     \caption{Distribution of Sound Event Count in Audio Prompts}
%     \label{fig:placeholder}
% \end{wrapfigure}

\begin{table}[t]
\centering
\caption{Demographic Information of Non-expert Annotators}
\label{tab:annotator_info}
\begin{tabular}{llllll}
\toprule
Annotator ID & Gender & Age & Academic Background & English Proficiency & Education Level \\
\midrule
ruike003 & Female & 26 & Network and New Media & CET-6 & Bachelor's Degree \\
ruike017 & Male   & 24 & Business English & CET-6 & Bachelor's Degree \\
ruike002 & Male   & 23 & Electrical Automation & Undergraduate-level & Bachelor's Degree \\
ruike001 & Female & 25 & Communication Engineering & Undergraduate-level & Bachelor's Degree \\
ruike020 & Male   & 25 & Interior Design & Undergraduate-level & Bachelor's Degree \\
ruike011 & Female & 23 & Advertising Design and Production & Undergraduate-level & Bachelor's Degree \\
ruike021 & Female & 20 & Business English & CET-4 & Bachelor's Degree \\
ruike012 & Female & 21 & Journalism & CET-6 & Bachelor's Degree \\
ruike016 & Male   & 26 & Robotics Industry & CET-4 & Bachelor's Degree \\
\bottomrule
\end{tabular}
\end{table}

\begin{table}[t]
\centering
\caption{Demographic Information of Expert Annotators}
\label{tab:expert_annotator_info}
\begin{tabular}{lllll}
\toprule
Annotator ID & Gender & Age & English Proficiency & Academic Background and Education \\
\midrule
yinyue01 & Female & 39 & CET-4 & M.S. in Music Performance, Shenyang Conservatory of Music \\
yinyue02 & Female & 40 & CET-4 & B.A. in Music Education, Hebei Normal University \\
yinyue03 & Female & 43 & CET-4 & Music Education Program, Xingtai University \\
\bottomrule
\end{tabular}
\end{table}

\subsection{Evaluation Dimensions Detail}
\label{app:scoring_guidelines}

% 标注细则，参考表格2-3列，简单融合一下，不用写具体的1-2 3-4
\begin{table*}[t]
\centering
\caption{Audio Evaluation Rubric and Scoring Criteria}
\label{tab:rubric}
\small
\begin{tabular}{p{4cm}p{1cm}p{6cm}p{4.5cm}}
\toprule
\textbf{Dimension} & \textbf{Score Range} & \textbf{Criteria} & \textbf{Examples} \\
\midrule
\textbf{Production Quality} & 1--10 & Evaluates audio clarity, fidelity, dynamic range, frequency response, and spatial characteristics. \textbf{1--2}: Severe distortion/noise/artifacts, audio elements completely indistinguishable. \textbf{3--4}: Noticeable distortion/noise, elements barely distinguishable. \textbf{5--6}: Partial distortion/noise present, elements vaguely distinguishable. \textbf{7--8}: Clear sound, no obvious distortion/noise, slight blurriness in low/high frequencies. \textbf{9--10}: Very clear sound, no distortion/noise, natural volume, balanced frequency, professional-grade recording/mixing/mastering. & Positive: \texttt{/positive/quality.wav} Negative: \texttt{/negative/quality.wav} \\
\midrule
\textbf{Production Complexity} & 1--10 & Assesses richness of audio elements, layering, and mixing complexity. \textbf{1--5}: Single audio element (e.g., solo speech, pure piano) or fewer than 3 elements. \textbf{6--10}: More than 3 audio elements (e.g., speech + music + sound effects, multi-part orchestral music). & Positive: \texttt{/positive/complexity.wav} Negative: \texttt{/negative/complexity.wav} \\
\midrule
\textbf{Content Enjoymen} & 1--10 & Evaluates emotional resonance, artistic skill, creativity, and personal experience. \textbf{1--5}: Fails to evoke emotion or resonance; poor performance/creation skill; lacks innovation and uniqueness; low overall preference. \textbf{6--10}: Evokes emotion or resonance; high performance/creation skill; demonstrates innovation and uniqueness; high overall preference. & - \\
\midrule
\textbf{Content Usefulness} & 1--10 & Measures audio quality as creative material and its applicability for high-quality content production. \textbf{1--2}: Unusable in any context. \textbf{3--4}: Cannot be used directly, requires extensive post-processing. \textbf{5--6}: Average quality, partially usable after minor post-processing. \textbf{7--8}: Good quality, suitable for amateur-level content (e.g., vlogs, short videos). \textbf{9--10}: High-quality material with appealing content, directly usable for professional production (e.g., film dubbing, broadcast-grade sound effects). & Positive: \texttt{/positive/utility.wav} Negative: \texttt{/negative/utility.wav} \\
\midrule
\textbf{Textual Alignment} & 1--10 & Assesses alignment between audio content and text, including accuracy, synchronization, and information completeness. \textbf{1--2}: Audio completely unrelated to text. \textbf{3--4}: Contains only minor elements from text, or includes substantial non-text content with inconsistent ordering. \textbf{5--6}: Contains partial text elements with some non-text content. \textbf{7--8}: Audio content matches text well with minor ordering inconsistencies. \textbf{9--10}: Audio content perfectly matches text with completely consistent ordering. & Positive: \texttt{/example/wav/S001\_P001} Negative: \texttt{/example/wav/S001\_P002} \\
\bottomrule
\end{tabular}
\end{table*}
The audio evaluation rubric adopted throughout this work is summarized in Table~\ref{tab:rubric}. The rubric decomposes audio assessment into five complementary dimensions covering both technical fidelity and content-level attributes. Production Quality and Production Complexity focus on signal-level characteristics and structural richness of the audio, respectively, capturing clarity, distortion, and compositional layering. Content Enjoyment reflects subjective listener preference, emotional engagement, and perceived creativity, while Content Usefulness evaluates the practicality of the audio as reusable material for downstream content creation and professional production. Finally, Textual Alignment measures the semantic and temporal correspondence between the audio and its associated text. Each dimension is rated on a unified 1–10 scale with clearly defined anchor descriptions to ensure consistency across evaluators. Representative positive and negative audio examples are provided to further standardize scoring criteria and reduce subjective variance.

%%%%%%%%%%%%%%%%%%%%%%%%%%%%%%%%%%%%%%%%%%%%%%%%%%%%%%%%%%%%%%%%%%%%%%%%%%%%%%%
%%%%%%%%%%%%%%%%%%%%%%%%%%%%%%%%%%%%%%%%%%%%%%%%%%%%%%%%%%%%%%%%%%%%%%%%%%%%%%%

\end{document}